\documentclass[showpacs,preprintnumbers,amsmath,amssymb,twocolumn]{revtex4}
\usepackage{amsmath}
\usepackage{graphicx}
\usepackage{epsf}
\usepackage{psfrag}
\usepackage{epsfig}
\usepackage{graphics}

\begin{document}
\newcommand{\e}{\mbox{e}}
\newcommand{\la}{\bar{\tilde{\lambda}}}
\newcommand{\be}{\begin{equation}}
\newcommand{\ee}{\end{equation}}
\newcommand{\bq}{\begin{eqnarray}}
\newcommand{\eq}{\end{eqnarray}}
\newcommand{\intx}{\int^{1}_{0} dx}
\newcommand{\pbruto}{\hbox{$p \!\!\!{\slash}$}}
\newcommand{\eps}{\epsilon}
\newcommand{\qbruto}{\hbox{$q \!\!\!{\slash}$}}
\newcommand{\lbruto}{\hbox{$l \!\!\!{\slash}$}}
\newcommand{\kbruto}{\hbox{$k \!\!\!{\slash}$}}

\def\cc{\c{c}}
\def\CC{\c{C}}
\def\ca{\c{c}\~{a}}
\def\co{\c{c}\~{o}}
\def\CA{\c{C}\~{A}}
\def\CO{\c{C}\~{O}}
\def\ii{\'{\i}}
\def\II{\'I}

\title{{\bf Ultraviolet and Infrared Divergences in Implicit Regularization: a Consistent Approach}}

\date{\today}

\author{H. G. Fargnoli$^{(a)}$} \email[]{hfargnoli@fisica.ufmg.br}
\author{A. P. Ba\^eta Scarpelli} \email []{scarp@fisica.ufmg.br}
\author{L. C. T. Brito$^{(b)}$} \email []{lctbrito@fisica.ufmg.br}
\author{B. Hiller$^{(c)}$} \email[]{brigitte@teor.fis.uc.pt}
\author{Marcos Sampaio$^{(a)}$} \email []{msampaio@fisica.ufmg.br}
\author{M. C. Nemes$^{(a)}$}\email[]{mcnemes@fisica.ufmg.br}
\author{A. A. Osipov$^{(c,d)}$} \email[]{osipov@nu.jinr.ru}

\affiliation{(a) Universidade Federal de Minas Gerais - Departamento de F\ii sica - ICEx \\ P.O. BOX 702, 30.161-970, Belo Horizonte MG -
Brazil}
\affiliation{(b) Universidade Federal de Lavras - Departamento de Ci\^encias Exatas\\
P. O. BOX 3037, 37.200-000, Lavras MG - Brazil}
\affiliation{(c) Departamento de Física, Centro de Física Computacional,
Faculdade de Ci\^encias e Tecnologia, Universidade de Coimbra,
P-3004 516 Coimbra - Portugal}
\affiliation{(d) Dzhelepov Laboratory of Nuclear Problems, JINR 141980 Dubna, Russia}
\begin{abstract}

\noindent
Implicit Regularization is a 4-dimensional regularization initially conceived to treat ultraviolet divergences. It has been successfully tested in several instances in the literature, more specifically in those where Dimensional Regularization does not apply. In the present contribution we extend the method to handle infrared divergences as well. We show that the essential steps which rendered Implicit Regularization adequate in the case of ultraviolet divergences have their counterpart for infrared ones. Moreover we show that a new scale appears, typically an infrared scale which is completely independent of the ultraviolet one. Examples are given.
\end{abstract}

\pacs{11.10.Gh, 11.15.Bt, 11.30.Qc}

\maketitle

\section{Introduction}

The Implicit Regularization technique is a scheme proposed in \cite{primeiro} with the main purpose to provide for a clear separation between the divergent content of physical amplitudes (regularization dependent) and its finite physical content (regularization independent).

Among the essential properties of this regularization which allowed for its success \cite{ri,ri2,ri3,ri4,rifim} some are worth stressing:\\
a) all calculations are performed in the dimension specific to a given theory (in the present work we deal with four dimensions), which opens the way to solve problems beyond the scope of Dimensional Regularization;\\
b) the technique we employ does not, in any step of the calculation, modify the original integrand. Using a mathematical algebraic relation we are able to separate divergent from finite parts in a clear way;\\
c) in what regards massless theories or even massive ones we are able to introduce an arbitrary scale which has been shown to play the role of a renormalization scale. Now a word about infrared divergences is in order. In cases when the theory is infrared safe such divergences will appear in the pertubative calculation within Implicit Regularization. We have shown in this case that an infrared regulator can be used to manipulate the integral and then an algebraic identity which replaces this infrared regulator by an arbitrary scale (the renormalization scale) restore the infrared safeness. In the case that genuine infrared divergences appear, this process cannot be applied and as we show in this contribution, an infrared scale emerges which is independent and can be completely distinguished from the ultraviolet one. This result is an important technical step in the direction of solving the NSVZ beta function problem \cite{Novikov, Murayama, Murayama2, Mas};\\
d) and last, but not least, we discuss symmetry preservation. As shown in \cite{zee} a diagrammatic proof to all orders of the preservation of gauge symmetry in QED is possible if and only if shifts in the integrations of internal loops are allowed. In several instances involving non-abelian gauge theories and supersymmetric ones we have also shown that symmetries will be preserved if and only if surface terms originated from shifts in perturbative integrals are all set to zero \cite{zumino, luellerson}.

Of course Implicit Regularization is not the only attempt throughout the years to construct a regularization which works in four dimensions. Given the spectacular success of Dimensional Regularization a natural extension has been tried: Dimensional Reduction. This generalization is known to work well at the one loop level. At two loops it has also been tried \cite{Stockinger} but its extension to higher loops do not preserve supersymmetry automatically. A successful theory in four dimensions has been constructed around the same time as Implicit Regularization: Differential Regularization \cite{DiffRen1, DiffRen2, DiffRen3, DiffRen4, DiffRen5, DiffRen6, DiffRen7, DiffRen8}. The main difference is that the latter works in configuration space while the former works directly in momentum space. Recently the equivalence of both at one loop level was shown \cite{carlos, carlos2}. Implicit Regularization has been consistently extended to n-loops in gauge theories \cite{Edson} while Differential Regularization has not yet.

The purpose of this paper is to show that infrared divergences as well as ultraviolet ones can be treated in a symmetrical way, which gives us further confidence in the method, in addition to an elegant formulation for massless theories.

\section{Formal Analogies in Handling Ultraviolet and Infrared Divergences}

The main ideia of Implicit Regularization is based on the recursively use within Feynman integrals \footnotemark[34]\footnotetext[34]{We assume that the integrals are regularized - by an implicit regularization - in such a way that formal manipulations of the integrand are possible.} of the algebraic identity
\begin{eqnarray}
\label{eq1}
&&\frac{1}{(p_{i}-k)^{2}-m^{2}}=\frac{1}{(k^{2}-m^{2})}\nonumber\\
&&-\frac{p_{i}^{2}-2p_{i}\cdot k}{(k^{2}-m^{2})[(p_{i}-k)^{2}-m^{2}]}
\end{eqnarray}
in order to isolate the divergent pieces as basic divergent integrals (BDI) which are independent of the external momenta $p_{i}$ and do not need to be evaluated.

Let us look at the logarithmically ultraviolet divergent integral:
\be\label{ultradiv}
I = \int\frac{d^{4}k}{(2\pi)^{4}}\frac{1}{(k^{2}-m^{2}+i\epsilon)((p-k)^{2}-m^{2}+i\epsilon)},
\ee
where $p$ stands for the external momentum and $m$ for the mass of the theory. In the following we drop the $i\epsilon$ in the propagators, but it is tacitly assumed that they depend on them. Applying (\ref{eq1}) once, we obtain
\be\label{eq2}
I = \int^{\Lambda}_{k}\frac{1}{(k^{2}-m^{2})^{2}}-\int^{\Lambda}_{k}\frac{p^{2}-2p\cdot k}{(k^{2}-m^{2})^{2}[(p-k)^{2}-
m^{2}]}.
\ee
Hereafter the superscript $\Lambda$ designates that the integral is regularized and $\int_{k}\equiv\int\frac{d^{4}k}{(2\pi)^{4}}$.

The first term on the right hand side is independent of the physical momentum and belongs to the class of BDI. This one is called $I_{log}(m^{2})$ and we leave it (and also the other BDI's) in this integral form in order to avoid explicit contact with specific regularizations. The second term is finite and can be evaluated in usual ways.

Note that the mass $m$ is still part of $I_{log}$. If the theory is massive, it may be left there and used as a renormalization group scale. However if the theory is massless, $m$ has a completely different meaning. It represents an infrared regulator which has to be set to zero in the end of the calculation. In this case it is mandatory to substitute the mass in $I_{log}$ by an arbitrary scale $\lambda$ which will play the role of the renormalization scale. For this purpose we make use of another mathematical identity:
\be\label{eq3}
I_{log}(m^{2})=I_{log}(\lambda^{2})+b\ln\left(\frac{\lambda^{2}}{m^{2}}\right),
\ee
where $b\equiv\frac{i}{(4\pi)^{2}}$.

Inserting this relation in (\ref{eq2}) we obtain
\be\label{ultravioleta1}
I = I_{log}(\lambda^{2})-b\ln\left(-\frac{p^{2}}{\lambda^{2}}\right)+2b,
\ee
in terms of the arbitrary parameter $\lambda$.
An important observation should be made at this point. The structure of the finite parts of (\ref{eq2}) involves logarithms, precisely what one needs to cancel the infrared regulator in favor of $\lambda$.
\vspace{0.5cm}

Now let us proceed with an integral which illustrates the appearance of an infrared BDI in Implicit Regularization:
\be\label{infra}
U = \int\frac{d^{4}k}{(2\pi)^{4}}\frac{1}{k^{4}(p-k)^{2}}.
\ee
By power counting $U$ is infrared divergent and ultraviolet finite. In order to be able to use all the mathematics developed for ultraviolet divergent integrals we firstly note the following:
\begin{eqnarray}
\label{kquatro}
&&\frac{1}{(k^{2}+i\epsilon)^2}=\nonumber\\
&&-\int^{\Lambda} d^{4}u\ e^{iku}\int^{\Lambda}\frac{d^{4}z}{(2\pi)^{4}}\frac{1}{(z^{2}-i\epsilon)((z-u)^{2}-i\epsilon)},\nonumber\\
\end{eqnarray}
where $z$ and $u$ are configuration variables, and use was made of the Fourier transforms from momentum to configuration space (appendix C). The explicit $i\epsilon$ dependence in configuration space will also be omitted in the following.

Note the striking similarity between the above integral in $z$ and equation (\ref{ultradiv}). We can thus write the result immediately (the opposite sign as compared to (\ref{ultravioleta1}) is due to the $i\epsilon$ prescriptions in configuration and momentum space)
\begin{eqnarray}
I(u^{2})&=&\lim_{\mu\rightarrow 0} \int^{\Lambda}_{z}\frac{1}{(z^{2}-\mu^{2})((z-u)^{2}-\mu^{2})}\nonumber\\
&=& -(\tilde{I}_{log}(\tilde{\lambda}^{-2})-b\ln\left(-u^{2}\tilde{\lambda}^{2}\right)+2b).\label{infrabdi}
\end{eqnarray}
But now $\tilde{I}_{log}(\tilde{\lambda}^{-2})$ is an infrared BDI and (\ref{eq3}) has been used in order to eliminate
$\mu^{2}$ in favor of the infrared scale $\tilde{\lambda}$.

Using this result and (\ref{kquatro}) in (\ref{infra}) we have
\begin{eqnarray}
\label{U}
&&U = -\int^{\Lambda}_{k}\frac{1}{(p-k)^{2}}\int^{\Lambda} d^{4}u\ e^{iku}I(u^{2})\nonumber\\
&&=-\frac{i}{(4\pi)^{2}}\int^{\Lambda}_{k}\int^{\Lambda} d^{4}u\int^{\Lambda} d^{4}x\frac{e^{i(p-k)x}}{x^{2}}e^{iku}I(u^{2})\nonumber\\
&&=\frac{i}{(4\pi)^{2}}\int^{\Lambda} d^{4}u\ \frac{e^{-ipu}}{u^{2}}\left(\tilde{I}_{log}(\tilde{\lambda}^{-2})\right.\nonumber\\
&&\left.-b\ln\left(-u^{2}\tilde{\lambda}^{2}\right)+2b\right)\nonumber\\
&&=\frac{1}{p^{2}}\left(\tilde{I}_{log}(\tilde{\lambda}^{-2})+b\ln\left(-\frac{p^{2}}{\bar{\tilde{\lambda}}^{2}}\right)+2b\right)\label{diviv}.
\end{eqnarray}
Here and henceforth the barred and the corresponding unbarred variables
are related as defined in appendix C.

Before closing this paragraph we emphasize that the extraction and classification of all infrared BDI follows the same strategy, of Fourier transforming the integrals, solving them in configuration space, and transforming the solution back to momentum space. As already remarked in \cite{Mas}, introducing the infrared scale in the Fourier transformed space has the convenient property of rendering this scale apriori independent from the ultraviolet scale.

\section{Infrared and Ultraviolet Divergent Integrals: Scale Separation}

Consider as a second example the following integral that is both ultraviolet and infrared divergent
\be
G(\bar{\tilde{\lambda}},\lambda)=\int_k\frac{1}{k^{4}}
\ln\left(\frac{-k^{2}}{\bar{\tilde{\lambda}}^{2}}\right)\ln^{2}\left(-\frac{k^{2}}{\lambda^{2}}\right).
\ee
Multiplying and dividing by $(p-k)^{2}$ allows us to separate
infrared and ultraviolet divergences, \be
G(\bar{\tilde{\lambda}},\lambda)=G_{uv}+p^{2}G_{ir}-2p^{\mu}G_{\mu,finite}
\ee where \be
G_{uv}=\int^{\Lambda}_k\frac{1}{k^{2}(p-k)^{2}}\ln\left(\frac{-k^{2}}
{\bar{\tilde{\lambda}}^{2}}\right)\ln^{2}\left(\frac{-k^{2}}{\lambda^{2}}\right),
\ee \be
G_{ir}=\int^{\Lambda}_k\frac{1}{k^{4}(p-k)^{2}}\ln\left(\frac{-k^{2}}
{\bar{\tilde{\lambda}}^{2}}\right)\ln^{2}\left(-\frac{k^{2}}{\lambda^{2}}\right)
\ee and \be
G_{\mu,finite}=\int^{\Lambda}_k\frac{k_{\mu}}{k^{4}(p-k)^{2}}\ln\left(\frac{-k^{2}}
{\bar{\tilde{\lambda}}^{2}}\right)\ln^{2}\left(\frac{-k^{2}}{\lambda^{2}}\right).
\ee
With Implicit Regularization rules it was shown in \cite{edsontese} (see also \cite{edson2}) that
\bq && \int^{\Lambda}_k
\frac{1}{k^{2}(p-k)^{2}}\ln^{n}\left(-\frac{k^{2}}{\lambda^{2}}\right)=\nonumber
\\&&I_{log}^{(n+1)}(\lambda^{2})+b\Bigg\{
n!-\sum_{i=0}^{n+1}(-1)^{n-i+1}\frac{n!}{i!}\times \nonumber
\\ && \ln^{i}\left(\frac{-p^{2}}{\lambda^{2}}\right)\Bigg\}, \eq
with $\lambda$ being an ultraviolet scale and $I_{log}^{(n+1)}(\lambda^{2})\equiv \int^{\Lambda}_{k}\frac{1}{(k^{2}-\lambda^{2})^{2}}\ln^{n}\left(\frac{-k^{2}}{\lambda^{2}}\right)$. Using this we
easily evaluate \bq &&
G_{uv}=\ln\left(\frac{\lambda^{2}}{\bar{\tilde{\lambda}}^{2}}\right)\Bigg\{
I_{log}^{(3)}(\lambda^{2})+b\Bigg[4-2\ln\left(\frac{-p^{2}}{\lambda^{2}}\right)+\nonumber
\\ && \ln^{2}\left(\frac{-p^{2}}{\lambda^{2}}\right)-
\frac{1}{3}\ln^{3}\left(\frac{-p^{2}}{\lambda^{2}}\right)\Bigg]\Bigg\}
\nonumber \\ &&
+I_{log}^{(4)}(\lambda^{2})+b\Bigg[6\ln\left(\frac{-p^{2}}{\lambda^{2}}\right)-
3\ln^{2}\left(\frac{-p^{2}}{\lambda^{2}}\right)+ \nonumber
\\  && \ln^{3}\left(\frac{-p^{2}}{\lambda^{2}}\right)-
\frac{1}{4}\ln^{4}\left(\frac{-p^{2}}{\lambda^{2}}\right)\Bigg].
\eq In order to evaluate $G_{ir}$ we write it as \bq &&
G_{ir}=\ln^{2}\left(\frac{\lambda^{2}}{\bar{\tilde{\lambda}}^{2}}\right)\int^{\Lambda}_k
\frac{1}{k^{4}(p-k)^{2}}\ln\left(\frac{-k^{2}}{\bar{\tilde{\lambda}}^{2}}\right)-
\nonumber \\ &&
2\ln\left(\frac{\lambda^{2}}{\bar{\tilde{\lambda}}^{2}}\right)\int^{\Lambda}_k
\frac{1}{k^{4}(p-k)^{2}}
\ln^2\left(\frac{-k^{2}}{\bar{\tilde{\lambda}}^{2}}\right) \nonumber \\
&&+\int^{\Lambda}_k \frac{1}{k^{4}(p-k)^{2}}
\ln^{3}\left(\frac{-k^{2}}{\bar{\tilde{\lambda}}^{2}}\right),\eq
composed of three terms with integrals in $k$ which for
definiteness we name $G_1$, $G_2$ and $G_3$. The integral
\be
G_{1}(p)=\int^{\Lambda}_{k}\frac{1}{k^{4}(p-k)^{2}}\ln\left(\frac{-k^{2}}{\bar{\tilde{\lambda}}^{2}}
\right) \ee can be rewritten with the aid of the Fourier transforms in appendix C as:
\be G_{1}(p) = \frac{i}{(4 \pi^2)^3}\int^{\Lambda} d^{4}x
\frac{e^{ipx}}{x^2} \int^{\Lambda} d^{4}z \frac{\ln(-z^2 \tilde{\lambda}^2)}{z^2(z+x)^2}.
\ee
The integral in $z$ is divergent at large distances (infrared) and finite at short distances. This integral is well known from two loop Feynman diagram calculations and was evaluated in momentum space \cite{Edson}. Here we just display the result translated to configuration space
\begin{equation*}
-(2\pi)^{4}\left[\tilde{I}_{log}^{(2)}(\tilde{\lambda}^{-2}) +b\ln\left(-x^{2}\tilde{\lambda}^{2}\right)-\frac{b}{2}\ln^{2}\left(-x^{2}\tilde{\lambda}^{2}\right)\right]
\end{equation*}
where $\tilde{I}_{log}^{(2)}(\tilde{\lambda}^{-2})\equiv \int^{\Lambda} \frac{d^{4}z}{(2\pi)^{4}} \frac{1}{(z^{2}-\tilde{\lambda}^{-2})^{2}}\ln\left(-z^{2}\tilde{\lambda}^{2}\right)$ is another infrared logarithmic BDI. Therefore integrating over $x$ leads to
\begin{eqnarray}
\label{G1}
G_{1}(p) &=&-\frac{1}{p^{2}}\tilde{I}^{(2)}_{log}(\tilde{\lambda}^{-2})+\frac{b}{p^{2}}\left\{
\frac{1}{2}\ln^{2}\left(\frac{-p^{2}}{\bar{\tilde{\lambda}}^{2}}\right)\right.\nonumber\\
&+&\left.\ln\left(\frac{-p^{2}}{\bar{\tilde{\lambda}}^{2}}\right)\right\},
\end{eqnarray}

Similarly  the Fourier tramsform of \be G_{2}(p)=\int^{\Lambda}_k
\frac{1}{k^{4}(p-k)^{2}}\ln^{2}\left(\frac{-k^{2}}
{\bar{\tilde{\lambda}}^{2}}\right)\ee reads \begin{eqnarray*} &&
\tilde{G_{2}}(x)=\frac{i}{(4\pi^{2})x^{2}}\Bigg\{
\tilde{I}_{log}^{(3)}(\tilde{\lambda}^{-2})+b\Bigg[4-2\ln\left(-x^{2}\tilde{\lambda}^{2}\right)\\
&&+
\ln^{2}\left(-x^{2}{\tilde{\lambda}}^{2}\right)-\frac{1}{3}\ln^{3}\left(-x^{2}{\tilde{\lambda}}^{2}
\right)\Bigg]\Bigg\}, \end{eqnarray*} so
\bq && G_{2}(p)
=\frac{1}{p^{2}}\tilde{I}_{log}^{(3)}({\tilde{\lambda}}^{-2})+\frac{b}{p^{2}}\Bigg\{
4+2\ln\left(\frac{-p^{2}}{\bar{\tilde{\lambda}}^{2}}\right)\nonumber
\\ && +\ln^{2}\left(\frac{-p^{2}}
{\bar{\tilde{\lambda}}^{2}}\right)+ \frac{1}{3}\left[\ln^{3}\left(\frac{-p^{2}}{\bar{\tilde{\lambda}}^{2}}\right)+
4\zeta(3)\right]\Bigg\}.\nonumber\\
&& \eq Finally the Fourier transform of \be G_{3}(p)=\int^{\Lambda}_k
\frac{1}{k^{4}(p-k)^{2}}\ln^{3}\left
(\frac{-k^{2}}{\bar{\tilde{\lambda}}^{2}}\right)\ee is
\bq &&
\tilde{G_{3}}(x)=-\frac{i}{(4\pi^{2})x^{2}}\Bigg\{
\tilde{I}_{log}^{(4)}({\tilde{\lambda}}^{-2})+b\Bigg[6\ln\left(-x^{2}{\tilde{\lambda}}^{2}\right)
\nonumber \\ &&
-3\ln^{2}\left(-x^{2}{\tilde{\lambda}}^{2}\right)+\ln^{3}\left(-x^{2}{\tilde{\lambda}}^{2}\right)-
\frac{1}{4}\ln^{4}\left(-x^{2}{\tilde{\lambda}}^{2}\right)\Bigg]\Bigg\}
\nonumber \\ && -\frac{i4\zeta(3)}{(4\pi^{2})x^{2}}\left\{
\tilde{I}_{log}({\tilde{\lambda}}^{-2})-b\ln\left(-x^{2}{\tilde{\lambda}}^{2}\right)+2b\right\}
\eq which back to momentum space yields

\bq &&G_{3}(p)=\frac{1}{p^{2}}\left[-\tilde{I}_{log}^{(4)}({\tilde{\lambda}}^{-2})-\tilde{I}_{log}({\tilde{\lambda}}^{-2})\zeta(3)\right]\nonumber\\
&&+ \frac{b}{p^{2}}\Bigg\{
-4\zeta(3)+6\ln\left(\frac{-p^2}{\bar{\tilde{\lambda}}^{2}}\right)
\nonumber \\
&&+3\ln^{2}\left(\frac{-p^{2}}{\bar{\tilde{\lambda}}^{2}}\right) +
\ln^{3}\left(\frac{-p^{2}}{\bar{\tilde{\lambda}}^{2}}\right)+\frac{1}{4}\ln^{4}
\left(\frac{-p^{2}}{\bar{\tilde{\lambda}}^{2}}\right)\Bigg\}.\nonumber\\
&&
\eq
Hence
\bq
G_{ir}=\ln^{2}\left(\frac{\lambda^{2}}{\bar{\tilde{\lambda}}^{2}}\right)
G_{1}-2\ln\left(\frac{\lambda^{2}}{\bar{\tilde{\lambda}}^{2}}\right)
G_{2} + G_{3}.
\eq

To evaluate $G_{\mu,finite}$ we need the following integrals which are straightforward to compute \cite{Edson}
\bq
&&\int_k\frac{k_{\mu}}{k^{4}(p-k)^{2}}\ln^{n}\left(\frac{-k^{2}}{\lambda^{2}}\right)=\nonumber\\
&&b\frac{p_{\mu}}{p^{2}} \sum_{\substack {j=0,\\P(j)=P(n)}}^{n}\frac{n!}{j!}\ln^{j}\left(\frac{-p^{2}}{\lambda^{2}}\right),
\eq
where the expression $P(j)=P(n)$ means $j$ with the same parity of $n$. Then \bq G_{\mu,finite}&=&
\ln\left(\frac{\lambda^{2}}{\bar{\tilde{\lambda}}^{2}}\right)\int^{\Lambda}_k
\frac{k_{\mu}}{k^{4}(p-k)^{2}}\ln^{2}\left(\frac{-k^{2}}{\lambda^{2}}\right)+\nonumber
\\ && \int^{\Lambda}_k\frac{k_{\mu}}{k^{4}(p-k)^{2}}
\ln^{3}\left(\frac{-k^{2}}{\lambda^{2}}\right)\eq is evaluated to
be \bq G_{\mu, finite}&=& b\frac{p_{\mu}}{p^{2}}\Bigg\{
\ln\left(\frac{\lambda^{2}}{\bar{\tilde{\lambda}}^{2}}\right)\left[2+\ln^{2}
\left(\frac{-p^{2}}{\lambda^{2}}\right)\right]\nonumber \\ &+&
6\ln\left(\frac{-p^{2}}{\lambda^{2}}\right) +
\ln^{3}\left(\frac{-p^{2}}{\lambda^{2}}\right)\Bigg\}. \eq
Therefore collecting just the finite parts we have \be G(\bar{\tilde{\lambda}},\lambda)=
G_{uv}+p^{2} G_{ir}-2p^{\mu}G_{\mu,finite} \nonumber
\ee \bq && = -b\Bigg\{
4\zeta(3)+2\left[1+\frac{4}{3}\zeta(3)\right]\ln\left(\frac{\lambda^{2}}{\bar{\tilde{\lambda}}^{2}}\right)\nonumber
\\ && +\ln^{2}\left(\frac{\lambda^{2}}{\bar{\tilde{\lambda}}^{2}}\right)-
\frac{1}{12}\ln^{4}\left(\frac{\lambda^{2}}{\bar{\tilde{\lambda}}^{2}}\right)\Bigg\}.
\eq

\section{Consistency in the Evaluation of Finite Integrals with Infrared and Ultraviolet Divergences in Intermediate Steps}

As a third example, let us consider the two loops massless master diagram in four dimensions $J(p)$:

\be
J(p)=\int_k\int_l
\frac{1}{k^{2}l^{2}\left(k-p\right)^{2}\left(k-l\right)^{2}\left(l-p\right)^{2}}.
\ee

$J(p)$ is finite and can be evaluated for instance using the Gegenbauer technique \cite{Rosner}. In appendix A it is shown within Implicit Regulazation that
\be\label{derivada} \int^{\Lambda}_{k,l}
\frac{\partial}{\partial k_{\mu}}\left\{ \frac{(k-l)_\mu}{k^{2}l^{2}\left(k-p\right)^{2}
\left(k-l\right)^{2}\left(l-p\right)^{2}}\right\} =0
\ee
which implies that
\bq
&& \int^{\Lambda}_{k,l}
\frac{1}{l^{4}\left(k-p\right)^{2}\left(k-l\right)^{2}\left(l-p\right)^{2}}= \nonumber\\
&& \int^{\Lambda}_{k,l} \frac{1}{k^{2}l^{4}\left(k-p\right)^{2}\left(l-p\right)^{2}}.
\eq

It is not difficult to show that, for $\alpha$ an arbitrary
positive constant,  \be \int^{\Lambda}_{k,l}
\partial^{\mu}\left\{ \frac{(k-l)_\mu \ln(-l^2/{\alpha^2})}{k^{2}l^{2}\left(k-p\right)^{2}
\left(k-l\right)^{2}\left(l-p\right)^{2}}\right\} =0,\ee in Implicit Regularization
with an additional logarithm \footnotemark[35]\footnotetext[35]{This identity is argued in
\cite{Smirnov} to be a four dimensional version of the integration
by parts method described in \cite{Chetyrkin1,Chetyrkin2}.}, which in turn
implies that \bq && J(p)= \int^{\Lambda}_k
\frac{1}{k^{2}(k-p)^{2}}\int^{\Lambda}_l
\frac{1}{\left(l-p\right)^{2}l^{4}}\ln\left(\frac{-l^{2}}{{\alpha}
^{2}}\right) \nonumber \\ && +
\int^{\Lambda}_k\frac{1}{k^{2}(k-p)^{2}}\int^{\Lambda}_l
\frac{1}{l^{2}\left(l-p\right)^{4}}\ln\left(\frac{-l^{2}}{{\alpha}
^{2}}\right)- \nonumber \\ && \int^{\Lambda}_{k,l}
\frac{1}{k^{2}l^{2}\left(k-l\right)^{2}
\left(l-p\right)^{4}}\ln\left(\frac{-l^{2}}{\alpha
^{2}}\right)-\nonumber
\\ && \int^{\Lambda}_{k,l}\frac{1}{(k-p)^{2}\left(k-l\right)^{2}\left(l-p\right)^{2}l^{4}}
\ln\left(\frac{-l^{2}}{{\alpha} ^{2}}\right) \nonumber \\
&& \equiv J_1(p) + J_2(p) - J_3(p) - J_4(p).\eq
Since $J(p)$ is finite, the divergences must cancel each other. This is demonstrated in appendix B showing the consistency of the method. So in the results below we will consider just the finite pieces of the integrals.

We start by evaluating
 \be J_{1}(p)= \int^{\Lambda}_k \frac{1}{k^{2}(k-p)^{2}} \int^{\Lambda}_l
\frac{1}{(l-p)^{2}l^{4}}
\ln\left(\frac{-l^{2}}{\alpha^{2}}\right).\ee

The integral in $l$, let us call it $I_{1}(p)$, is infrared
divergent,
\[ I_{1}(p)= \int^{\Lambda}_l
\frac{1}{(l-p)^{2}l^{4}}\ln\left(\frac{-l^{2}}{\alpha^{2}}\frac{\bar{\tilde{\lambda}}^{2}}{\bar{\tilde{\lambda}}^{2}}\right).\]
The infrared scale $\bar{\tilde{\lambda}}^{2}$ was introduced inside the logarithm to simplify the calculation separating $I_{1}(p)$ into two pieces:
\begin{eqnarray}\label{i1}
I_{1}(p)&=& \ln\left(\frac{\bar{\tilde{\lambda}}^{2}}{\alpha^{2}}\right)\int^{\Lambda}_l
\frac{1}{(l-p)^{2}l^{4}}\nonumber\\
&+& \int^{\Lambda}_l \frac{1}{(l-p)^{2}l^{4}}\ln\left(\frac{-l^{2}}{\bar{\tilde{\lambda}}^{2}}\right).
\end{eqnarray}
Both integrals have been already calculated, the first is proportional to $U$, eq. \ref{U}, the second is $G_1(p)$, \ref{G1}.

Finally putting the ultraviolet result (\ref{ultravioleta1}) together with $I_{1}(p)$ we get
\begin{eqnarray}
&&J_{1}(p) = \frac{b^{2}}{p^{2}}\left[4\ln\left(\frac{\la^{2}}{\alpha^{2}}\right)+2\ln\left(\frac{\la^{2}}{\alpha^{2}}\right)\ln\left(\frac{\lambda^{2}}{\la^{2}}\right)\right.\nonumber\\
&&+\ln\left(\frac{\lambda^{2}}{\la^{2}}\right)\ln\left(\frac{-p^{2}}{\la^{2}}\right)-\frac{1}{2}\ln\left(\frac{-p^{2}}{\lambda^{2}}\right)\ln^{2}\left(\frac{-p^{2}}{\la^{2}}\right)\nonumber\\
&&\left.+2\ln\left(\frac{-p^{2}}{\la^{2}}\right)-\ln\left(\frac{-p^{2}}{\lambda^{2}}\right)\ln\left(\frac{\la^{2}}{\alpha^{2}}\right)\ln\left(\frac{-p^{2}}{\la^{2}}\right)\right].\nonumber\\
&&
\end{eqnarray}

The other three integrals $J_{2}(p)$, $J_{3}(p)$ and $J_{4}(p)$ are evaluated in a similar fashion \footnotemark[36]\footnotetext[36]{The $J_{4}(p)$ calculation is easier if we first take a Fourier transform, evaluate $J_{4}(x)$ and then take the backward Fourier transform. The same thing can be done with $J_{1}(p)$, $J_{2}(p)$ and $J_{3}(p)$.}. We only quote the results:
\begin{eqnarray}
&&J_{2}(p) = \frac{b^{2}}{p^{2}}\left[2\ln\left(\frac{\lambda^{2}}{\la^{2}}\right)\ln\left(\frac{-p^{2}}{\alpha^{2}}\right)+4\ln\left(\frac{-p^{2}}{\alpha^{2}}\right)\right.\nonumber\\
&&\left.-\ln\left(\frac{-p^{2}}{\lambda^{2}}\right)\ln\left(\frac{-p^{2}}{\alpha^{2}}\right)\ln\left(\frac{-p^{2}}{\la^{2}}\right)\right],
\end{eqnarray}

\begin{eqnarray}
&&J_{3}(p) = \frac{b^{2}}{p^{2}}\left[2\ln\left(\frac{\lambda^{2}}{\la^{2}}\right)\ln\left(\frac{-p^{2}}{\alpha^{2}}\right)+4\ln\left(\frac{-p^{2}}{\alpha^{2}}\right)\right.\nonumber\\
&&\left.-\ln\left(\frac{-p^{2}}{\lambda^{2}}\right)\ln\left(\frac{-p^{2}}{\alpha^{2}}\right)\ln\left(\frac{-p^{2}}{\la^{2}}\right)-4\zeta(3)\right],
\end{eqnarray}

\begin{eqnarray}
&&J_{4}(p) = \frac{b^{2}}{p^{2}}\left[4\ln\left(\frac{\la^{2}}{\alpha^{2}}\right)+2\ln\left(\frac{\la^{2}}{\alpha^{2}}\right)\ln\left(\frac{\lambda^{2}}{\la^{2}}\right)\right.\nonumber\\
&&+\ln\left(\frac{\lambda^{2}}{\la^{2}}\right)\ln\left(\frac{-p^{2}}{\la^{2}}\right)-\frac{1}{2}\ln\left(\frac{-p^{2}}{\lambda^{2}}\right)\ln^{2}\left(\frac{-p^{2}}{\la^{2}}\right)\nonumber\\
&&+2\ln\left(\frac{-p^{2}}{\la^{2}}\right)-\ln\left(\frac{-p^{2}}{\lambda^{2}}\right)\ln\left(\frac{\la^{2}}{\alpha^{2}}\right)\ln\left(\frac{-p^{2}}{\la^{2}}\right)\nonumber\\
&&\left.-2\zeta(3)\right].
\end{eqnarray}

Finally $J_1(p) + J_2(p)-J_3(p)-J_4(p) = J(p)$ leading to the result
\be J(p)
= - \frac{6 \zeta(3)}{(4 \pi)^4 p^{2}},
\ee
in agreement with the literature \cite{Rosner}, \cite{Smirnov}.

It is important to say that the operations $R$ and $\tilde{R}$ that subtract ultraviolet and infrared divergences respectively \cite{Chetyrkin1,Chetyrkin2} could be defined as is done in other regularization schemes and could be used in the two examples above. However it is straightforward to see what these operations do within Implicit Regularization. They just remove the ultraviolet and infrared BDI's.

\section{Concluding Remarks}

In the present contribution we have shown how the important steps which led to the construction of Implicit Regularization, initially used only in ultraviolet divergent integrals, can be adapted to handle infrared divergent integrals with the same degree of consistency. We have shown that an infrared scale appears which is completely independent of the ultraviolet one. Moreover we show the consistency of the procedure for integrals involving both divergences, in several testing non trivial examples. The results presented here, albeit being formal, are essential to answer several open questions in the literature. A crucial step in this direction is that Implicit Regularization allows to trace the infrared or ultraviolet origin of a basic logarithmic divergence and to associate to each kind of divergence independent scale relations. The independence of these scales is an important technical issue, the lack of which has prevented finding solutions to several important problems such as the NSVZ beta function. This work provides for an adequate framework within which this problem may be handled.

\section{Appendix A - Verification of (\ref{derivada})}

Differentiating (\ref{derivada}) we have:
\begin{eqnarray}
&&A = \frac{\partial}{\partial k_{\mu}}\left(\frac{(k-l)_{\mu}}{D}\right) = \frac{1}{D}\left(2 + \frac{l^{2}}{k^{2}}+\frac{(p-l)^{2}}{(p-k)^{2}}\right.\nonumber\\
&&\left.-\frac{(k-l)^{2}}{k^{2}}-\frac{(k-l)^{2}}{(p-k)^{2}}\right),
\end{eqnarray}
with
\begin{equation*}
D = k^{2}l^{2}(k-l)^{2}(p-k)^{2}(p-l)^{2}.
\end{equation*}

Shifting the variables we get
\begin{equation}
A = \int^{\Lambda}_{k,l}\frac{1}{k^{4}(p-k)^{2}l^{2}(p-l)^{2}}-\int^{\Lambda}_{k,l}\frac{1}{k^{2}(p-k)^{4}l^{2}(l-k)^{2}}
\end{equation}

Using (\ref{ultravioleta1})
\begin{eqnarray}
&& A = \left(I_{log}(\lambda^{2})+2b\right)\int_{k}^{\Lambda}\frac{1}{k^{4}(p-k)^{2}}\nonumber\\
&&-b\ln\left(\frac{-p^{2}}{\lambda^{2}}\right)\int_{k}^{\Lambda}\frac{1}{k^{4}(p-k)^{2}}\nonumber\\
&&-\left(I_{log}(\lambda^{2})+2b\right)\int_{k}^{\Lambda}\frac{1}{k^{2}(p-k)^{4}}\nonumber\\
&&+ b\int^{\Lambda}_{k}\frac{1}{k^{2}(p-k)^{4}}\ln\left(\frac{-k^{2}}{\lambda^{2}}\right)
\end{eqnarray}

The last integral is evaluated using the procedure described in the main text and yields
\begin{eqnarray}
&&\frac{1}{p^{2}}\left[\tilde{I}_{log}(\tilde{\lambda}^{-2})\ln\left(\frac{-p^{2}}{\lambda^{2}}\right)+2b\ln\left(\frac{-p^{2}}{\lambda^{2}}\right)\right.\nonumber\\
&&\left.+b\ln\left(\frac{-p^{2}}{\lambda^{2}}\right)\ln\left(\frac{-p^{2}}{\la^{2}}\right)\right]
\end{eqnarray}

Upon insertion of (\ref{diviv}) all terms cancel and $A = 0$.

\section{Appendix B - Cancelation of divergent pieces}

Here we collect the divergent pieces of $J_{1}(p)$, $J_{2}(p)$, $J_{3}(p)$ and $J_{4}(p)$. Putting together the results (\ref{ultravioleta1}), (\ref{diviv}) and (\ref{G1}) we have:
\begin{eqnarray}
&&J_{1}(p) = \frac{1}{p^{2}}\left[I_{log}(\lambda^{2})-b\ln\left(\frac{-p^{2}}{\lambda^{2}}\right)+2b\right]\times\nonumber\\
&&\left[-\tilde{I}^{(2)}_{log}(\tilde{\lambda}^{-2})+b\ln\left(\frac{-p^{2}}{\bar{\tilde{\lambda}}^{2}}\right)+\frac{b}{2}\ln^{2}\left(\frac{-p^{2}}{\bar{\tilde{\lambda}}^{2}}\right)+\right.\nonumber\\ &&\left.\ln\left(\frac{\bar{\tilde{\lambda}}^{2}}{\alpha^{2}}\right)\left(\tilde{I}_{log}(\tilde{\lambda}^{-2})+b\ln\left(\frac{-p^{2}}{\bar{\tilde{\lambda}}^{2}}\right)+2b\right)\right]
\end{eqnarray}

So
\begin{eqnarray}
&&J_{1}^{div}(p)=\frac{1}{p^{2}}\left\{I_{log}(\lambda^{2})\left[-\tilde{I}^{(2)}_{log}(\tilde{\lambda}^{-2})+b\ln\left(\frac{-p^{2}}{\bar{\tilde{\lambda}}^{2}}\right)\right.\right.\nonumber\\
&&+\frac{b}{2}\ln^{2}\left(\frac{-p^{2}}{\bar{\tilde{\lambda}}^{2}}\right)+\ln\left(\frac{\bar{\tilde{\lambda}}^{2}}{\alpha^{2}}\right)\left(\tilde{I}_{log}(\tilde{\lambda}^{-2})\right.\nonumber\\
&&\left.\left.+b\ln\left(\frac{-p^{2}}{\bar{\tilde{\lambda}}^{2}}\right)+2b\right)\right]+\left(\tilde{I}_{log}(\tilde{\lambda}^{-2})\ln\left(\frac{\bar{\tilde{\lambda}}^{2}}{\alpha^{2}}\right)\right.\nonumber\\
&&\left.\left.-\tilde{I}^{(2)}_{log}(\tilde{\lambda}^{-2})\right)\left(2b-b\ln\left(\frac{-p^{2}}{\lambda^{2}}\right)\right)\right\}
\end{eqnarray}

The other divergent pieces are
\begin{eqnarray}
&&J^{div}_{2}(p) = \frac{1}{p^{2}}\left\{I_{log}(\lambda^{2})\left[\tilde{I}_{log}(\tilde{\lambda}^{-2})\ln\left(\frac{-p^{2}}{\alpha^{2}}\right)\right.\right.\nonumber\\
&&\left.+2b\ln\left(\frac{-p^{2}}{\alpha^{2}}\right)+b\ln\left(\frac{-p^{2}}{\alpha^{2}}\right)\ln\left(\frac{-p^{2}}{\la^{2}}\right)\right]\nonumber\\
&&\left.+\tilde{I}_{log}(\tilde{\lambda}^{-2})\ln\left(\frac{-p^{2}}{\alpha^{2}}\right)\left(2b - b\ln\left(\frac{-p^{2}}{\lambda^{2}}\right)\right)\right\}\nonumber\\
&&
\end{eqnarray}
\begin{eqnarray}
&&J^{div}_{3}(p) = \frac{1}{p^{2}}\left\{I_{log}(\lambda^{2})\left[\tilde{I}_{log}(\tilde{\lambda}^{-2})\ln\left(\frac{-p^{2}}{\alpha^{2}}\right)\right.\right.\nonumber\\
&&\left.+2b\ln\left(\frac{-p^{2}}{\alpha^{2}}\right)+b\ln\left(\frac{-p^{2}}{\alpha^{2}}\right)\ln\left(\frac{-p^{2}}{\la^{2}}\right)\right]\nonumber\\
&&+\tilde{I}_{log}(\tilde{\lambda}^{-2})\ln\left(\frac{-p^{2}}{\alpha^{2}}\right)\left(2b-b\ln\left(\frac{\alpha^{2}}{\lambda^{2}}\right)\right)\nonumber\\
&&\left.-\tilde{I}_{log}(\tilde{\lambda}^{-2})\ln^{2}\left(\frac{-p^{2}}{\alpha^{2}}\right)\right\}
\end{eqnarray}
\begin{eqnarray}
&&J^{div}_{4}(p) = \frac{1}{p^{2}}\left\{I_{log}(\lambda^{2})\left(-\tilde{I}_{log}^{(2)}(\tilde{\lambda}^{-2})\right.\right.\nonumber\\
&&+\frac{b}{2}\ln^{2}\left(\frac{-p^{2}}{\la^{2}}\right)+b\ln\left(\frac{-p^{2}}{\la^{2}}\right)+\ln\left(\frac{\la^{2}}{\alpha^{2}}\right)\tilde{I}_{log}(\tilde{\lambda}^{-2})\nonumber\\ &&\left.+b\ln\left(\frac{\la^{2}}{\alpha^{2}}\right)\ln\left(\frac{-p^{2}}{\la^{2}}\right)+2b\ln\left(\frac{\la^{2}}{\alpha^{2}}\right)\right)\nonumber\\
&&\left.+\tilde{I}_{log}^{(2)}(\tilde{\lambda}^{-2})\left(-1+\ln\left(\frac{\la^{2}}{\alpha^{2}}\right)\right)\left(2b-b\ln\left(\frac{-p^{2}}{\lambda^{2}}\right)\right)\right\}\nonumber\\
&&
\end{eqnarray}

Using these results in $J^{div}(p) = J^{div}_1(p) + J^{div}_2(p) - J^{div}_3(p) - J^{div}_4(p)$ it is easy to see the cancellation of the divergent pieces.

\section{Appendix C - Fourier Transforms}

The Fourier transform of any power of logarithmic functions  \cite{DiffRen1} can be obtained through the Fourier transform of power functions. Our results are given in Minkowski space, for which we use \cite{Grozin}:
\be\label{apendice1}
\int d^{4}x\ e^{ipx} \frac{1}{x^{2}}(-B^{2}x^{2})^{a} = -i\frac{4\pi^{2}}{p^{2}}\left(\frac{-4B^{2}}{p^{2}}\right)^{a}\frac{\Gamma(1+a)}{\Gamma(1-a)},
\ee
with $B$ arbitrary. Employing the relation \cite{kazakov}
\be
\frac{\Gamma(1+a)}{\Gamma(1-a)} = e^{-2\gamma a}exp\left[2\sum_{n=1}^{\infty}\frac{\zeta(2n+1)}{2n+1}a^{2n+1}\right],
\ee
where $\gamma = 0,5772...$ is the Euler-Mascheroni constant and $\zeta$ is the zeta Riemann function, and expanding $(-B^{2}x^{2})^{a}$ as a power series
\be
(-B^{2}x^{2})^{a} = \sum_{n=0}^{\infty}\frac{a^{n}}{n!}\ln^{n}\left(-x^{2}B^{2}\right)
\ee
one can compare both sides of (\ref{apendice1}). After some algebra one gets
$$ \sum_{n=0}^{\infty}\frac{a^{n}}{n!}\int
d^{4}x\ e^{ipx}\frac{1}{x^{2}}\ln^{n}\left(-x^{2}{B}^{2}\right)=
-i\frac{4\pi^{2}}{p^{2}} \times $$ \be
\exp\left[-a\ln\left(-\frac{p^{2}e^{2\gamma}}{4B^{2}}\right)-
2\sum_{n=1}^{\infty}\frac{\zeta(2n+1)}{2n+1}a^{2n+1}\right].\ee

Expanding the exponential in power series and equating the coefficients of equal powers of $a$ one obtains
\be \int d^{4}x\ \frac{e^{i p x}}{x^2} = - i\frac{4 \pi^2}{p^2}. \ee
\be \int d^{4}x\ \frac{e^{i p x}}{x^2}\ln (-x^2 B^2) =
i\frac{4 \pi^2}{p^2} \ln \Big(\frac{-p^2}{\bar{B}^2} \Big).
\ee \be \int d^{4}x\ \frac{e^{i p x}}{x^2}\ln^2 (-x^2 B^2) = -
i\frac{4 \pi^2}{p^2} \ln^2 \Big(\frac{-p^2}{\bar{B}^2} \Big).
\ee \be \int d^{4}x\ \frac{e^{i p x}}{x^2}\ln^3 (-x^2 B^2) =
i\frac{4 \pi^2}{p^2} \Bigg(\ln^3 \Big(\frac{-p^2}{\bar{B}^2}
\Big)+ 4 \zeta (3) \Bigg). \ee

\begin{eqnarray} \int d^{4}x\ \frac{e^{i p x}}{x^2}\ln^4 (-x^2 B^2) &=&
-i\frac{4 \pi^2}{p^2} \Bigg(\ln^4 \Big(\frac{-p^2}{\bar{B}^2}
\Big)\nonumber\\
&+& 16 \zeta(3)\ln \Big(\frac{-p^2}{\bar{B}^2}
\Big) \Bigg).\nonumber\\
&& \end{eqnarray}
with $\bar{B}^2 \equiv \frac{4}{e^{2\gamma}}B^{2}$.

\begin{center}\textbf{\small{ACKNOWLEDGMENTS}}\end{center}

This work was supported by CAPES and CNPq, by grants of FCT, FEDER, OE, POCI 2010, CERN/FP/83510/2008, and by HadronPhysics2, Grant Agreement n. 227431 under the Seventh Framework Programme of EU.


\begin{thebibliography}{99}

\bibitem{primeiro} O. A. Battistel, A. L. Mota and M. C. Nemes, \textit{Mod. Phys. Lett. \textbf{A 13}}, 1597 (1998).

\bibitem{ri} A. Brizola, O. A. Battistel, M. Sampaio and M. C. Nemes, \textit{Mod. Phys. Lett. \textbf{A 14}}, 1509 (1999).

\bibitem{ri2} O. A. Battistel and M. C. Nemes, \textit{Phys. Rev. \textbf{D 59}}, 055010 (1999).

\bibitem{ri3} A. P. Ba\^eta Scarpelli, M. Sampaio and M. C. Nemes, \textit{Phys. Rev. \textbf{D 63}}, 046004 (2001).

\bibitem{ri4} A. P. Ba\^eta Scarpelli, M. Sampaio, M. C. Nemes and B. Hiller, \textit{Phys. Rev. \textbf{D 64}}, 046013 (2001).

\bibitem{rifim} L. A. M. Souza, Marcos Sampaio, M. C. Nemes, \textit{Phys. Lett.} \textbf{B 632} 717 (2006).

\bibitem{Novikov} V. Novikov, M. Shifman, A. Vainstein, V.
Zakharov, \textit{Phys. Lett.} \textbf{\textit{B} 166} 329 (1985).

\bibitem{Murayama} N. A-Hamed and H. Murayama, \textit{Phys. Rev.} \textbf{\textit{D} 57} (1998)
6638.

\bibitem{Murayama2} N. A-Hamed and H. Murayama, \textit{JHEP} \textbf{0006} 030 (2000).

\bibitem{Mas} J. Mas, M. Perez-Victoria, C. Seijas, \textit{JHEP} \textbf{0203} 049 (2002).

\bibitem{zee} A. Zee, \textit{Quantum Field Theory in a Nutshell} (2003).

\bibitem{zumino}  D. Carneiro, A. P. Ba\^eta Scarpelli, M. Sampaio and M. C.
Nemes, \textit{JHEP} \textbf{12} 044 (2003).

\bibitem {luellerson} L. C. Ferreira, M. C. Nemes, M. Sampaio, \textit{Work in progress}.

\bibitem{Stockinger} D. Stockinger, \textit{Nucl. Phys. Suppl.} \textbf{160} (2006)
250; \textit{JHEP} \textbf{0503} 076 (2005).

\bibitem{DiffRen1} D. Z. Freedman, K. Johnson and J. I. Latorre, \textit{{Nucl. Phys.
\textbf{B}}} \textbf{371}, 353 (1992).

\bibitem{DiffRen2}P. E. Haagensen, \textit {{
Mod. Phys. Lett. \textbf{A}}} \textbf{7}, 893 (1992).

\bibitem{DiffRen3} R.
Mu\~noz-Tapia, \textit {{ Phys. Lett. \textbf{B}}} \textbf{295}, 95
(1992).

\bibitem{DiffRen4} D. Z Freedman, G. Grignani, K. Johnson and N. Rius,
\textit {{ Ann. Phys. }} \textbf{218}, 75 (1992).

\bibitem{DiffRen5} P. E. Haagensen,
J. I. Latorre, \textit {{ Ann. Phys.}} \textbf{221}, 77 (1993).

\bibitem{DiffRen6} G. Dunne, N. Rius, \textit {{Phys. Lett. \textbf{B}}}
\textbf{293}, 367 (1992).

\bibitem{DiffRen7} D. Z. Freedman, K.
Johnson, R. Mu\~noz-Tapia and X. Vilasis-Cardona, \textit {{Nucl.
Phys. \textbf{B}}} \textbf{395}, 454 (1993).

\bibitem{DiffRen8} P. E. Haagensen, J. I.
Latorre, \textit {{Phys. Lett. \textbf{B}}} \textbf{283}, 293
(1992).

\bibitem{carlos} C. R. Pontes, A. P. Ba\^eta Scarpelli, Marcos Sampaio, J. L.
Acebal, M. C. Nemes, \textit{Eur. Phys. J. C} \textbf{53} 121 (2008).

\bibitem{carlos2} Marcos Sampaio, A. P. Ba\^eta Sacarpelli, B. Hiller, A. Brizola, M.C. Nemes, S. Gobira, \textit{Phys. Rev.} \textbf{D} \textbf{65} 667 (2002) 125023.

\bibitem{Edson} E. W. Dias, A. P. Ba\^eta Scarpelli, L. C. T. Brito, Marcos Sampaio, M. C. Nemes,
\textit{Eur. Phys. J. } \textbf{C} \textbf{55} 667 (2008).

\bibitem{edsontese} E. W. Dias, Ph.D. thesis, Federal University of Minas Gerais, Brazil (2008).

\bibitem{edson2} E. W. Dias, A. P. Ba\^eta Scarpelli, L. C. T. Brito, H. G. Fargnoli, \textit{Braz.J.Phys.} \textbf{40} 228 (2010).

\bibitem{Rosner} J. L. Rosner, \textit{Ann. Phys.} \textbf{44} 11 (1967).

\bibitem{Smirnov} V. A. Smirnov, \textit{Theor. Math. Phys.} \textbf{108} 953 (1997).

\bibitem{Chetyrkin1} K. G. Chetyrkin and F. V. Tkachov, \textit{Nuc. Phys.}
\textbf{\textit{B} 192} 159 (1981).

\bibitem{Chetyrkin2} K. G. Chetyrkin and V. A. Smirnov, \textit{Phys. Lett.} \textbf{\textit{B} 144} 419 (1984).

\bibitem{Grozin} A. G. Grozin, \textit{Lectures on QED and QCD Practical Calculation and Renormalization of One- and Multi-Loop Feynman Diagrams} (2007).

\bibitem{kazakov} L. V. Avdeev, D. I. Kazakov, I. N. Kondrashuk, \textit{Int. J. Mod. Phys. \textbf{A 9}} 1067 (1994).

\end{thebibliography}
\end{document}